\documentclass[pra,superscriptaddress,showpacs
,preprint
]{revtex4}
\usepackage{amsmath,amssymb,amsfonts,graphicx}
\newcommand{\ket}[1]{\left|#1\right>}
\newcommand{\bra}[1]{\left<#1\right|}
\newcommand{\sket}[2]{\left|#1\right>\!{}_{{}_{\scriptstyle{#2}}}}
\newcommand{\sbra}[2]{{}_{{}_{\scriptstyle{#2}}}\!\left<#1\right|}
\newcommand{\ssket}[2]{\left|#1\right>\!{}_{{}_{#2}}}
\newcommand{\ssbra}[2]{{}_{{}_{#2}}\!\left<#1\right|}

\begin{document}

\title{Asymmetric universal entangling machine}
\author{D. B. Horoshko}
\email[E-mail: ]{dhoroshko@yahoo.com}
\affiliation{Laboratoire
PhLAM, Universit\'e de Lille 1, 59655 Villeneuve d'Ascq, France}
\affiliation{B.I.Stepanov Institute of Physics, Belarus National
Academy of Sciences, Minsk 220072 Belarus}
\author{M. I. Kolobov}
\affiliation{Laboratoire PhLAM, Universit\'e de Lille 1, 59655
Villeneuve d'Ascq, France}
\author{S. Ya. Kilin}
\affiliation{B.I.Stepanov Institute of Physics, Belarus National
Academy of Sciences, Minsk 220072 Belarus}
\date{\today}
\begin{abstract}
We give a definition of asymmetric universal entangling machine
which entangles a system in an unknown state to a specially
prepared ancilla. The machine produces a fixed state-independent
amount of entanglement in exchange to a fixed degradation of the
system state fidelity. We describe explicitly such a machine for
any quantum system having $d$ levels and prove its optimality. We
show that a $d^2$-dimensional ancilla is sufficient for reaching
optimality. The introduced machine is a generalization to a number
of widely investigated universal quantum devices such as the
symmetric and asymmetric quantum cloners, the symmetric quantum
entangler, the quantum information distributor and the
universal-NOT gate.
\end{abstract}
\maketitle

\section{Introduction}
Entanglement is a fundamental notion of quantum theory, having no
classical counterpart. A pure state of two quantum systems is
called entangled if it is not a product of a state for one system
and a state for another. This definition is similar to the
definition of statistical dependence in the classical theory.
However, entanglement implies stronger correlations between the
observables of both systems than it is allowed by the classical
statistical theory.

The first machine generating entanglement was described by
Schr\"odinger in 1935 in the very paper introducing the notion of
entanglement \cite{Schroedinger35}. In his famous
\textit{gedankenexperiment} Schr\"odinger describes a machine
which produces entanglement between the states of a cat and a
decaying nucleus placed inside a closed box together with a
diabolic machine, which kills the cat if the nucleus decays. The
nucleus is initially in a non-decayed state $\left|2\right>$ and
in one hour it has a probability of $\frac12$ to decay to the
state $\left|1\right>$ . If we denote the state of alive and dead
cat $\left|alive\right>$ and $\left|dead\right>$ respectively, and
the initial and final states of the diabolic machine
$\left|i\right>$ and $\left|f\right>$ respectively, then in one
hour the state of nucleus, diabolic machine and cat transforms (up
to normalization) like
\begin{equation}
\ket{2}\ket{i}\ket{alive}\longrightarrow\ket{2}\ket{i}\ket{alive}+\ket{1}\ket{f}\ket{dead}.
\label{1}
\end{equation}
The state in the right hand side of Eq. (\ref{1}) is entangled
since it is pure and is not a product. We see that the nucleus
together with the diabolic machine act as ``entangling machine''
for the cat, because they produce entanglement between themselves
and the cat. However, this machine works only if the initial state
of the cat is ``alive''. Indeed, if we place a dead cat inside the
box, then the transformation (under the natural assumption that
the diabolic machine does not affect the dead cat) will be
\begin{equation}
\ket{2}\ket{i}\ket{dead}\longrightarrow\ket{2}\ket{i}\ket{dead}+\ket{1}\ket{f}\ket{dead},
\label{2}
\end{equation}
and the states of nucleus and cat are not entangled. Thus we have
found that the entangling machine proposed by Schr\"odinger is
state-dependent, i.e. it produces entanglement for some input
states and does not for others.

All quantum devices used today for producing entanglement between
quantum systems share this feature with Schr\"odinger's machine:
they all require the quantum systems to be prepared in definite
initial states. These states can be quite ``natural'', e.g. the
vacuum state of two sub-harmonics in the process of parametric
down-conversion \cite{QI}. Nevertheless, it is interesting from
both fundamental and practical points of view to know if it is
possible to design a state-independent entangling machine which
generates equal non-zero amount of entanglement for any state of
the incoming system. Such a machine would have fundamental
importance because it would clarify the relation between the
notions of ``quantum state'' and ``quantum entanglement'', i.e. if
the local and non-local properties of quantum systems can be
manipulated independently from one another. From the practical
point of view such machine would have applications to preparation
of entangled state of two systems in the case where the initial
state of one of them is not fully known at the moment of
entangling interaction. This situation is typical for
eavesdropping in quantum cryptography and can be met in quantum
computation \cite{QI}.

Quantum machines generating entanglement in a universal (input
state-independent) way have been considered by Alber \cite{Alber}
and Bu\u zek and Hillery \cite{SymmEnt}. Both approaches aim at
entangling two similar systems (having the same number of levels)
in either antisymmetric \cite{Alber} or symmetric \cite{SymmEnt}
way. The latter type of entangling interaction has been recently
realized in an experiment \cite{SymmEntExper}. In the present
paper we develop an alternative approach which does not require
two systems to be similar to one another (cf. cat and nucleus
above). As a consequence, our approach cannot and does not impose
any symmetry condition on the output, and the machine of our
interest can be called asymmetric universal entangling machine
(AUEM). As we will see later, to obtain a non-trivial machine we
will need to impose an additional constraint, linking the machine
output to its input. Other and deeper differences between the
approach of the present paper and that of Refs. \cite{Alber} and
\cite{SymmEnt} will be clarified in the next Sections.

The paper is structured as follows. In Sec. II we give a
definition of asymmetric universal entangling machine. In Sec. III
we deduce the explicit form of a transformation entangling a
$d$-level system in an a priori unknown state to a $d^2$-level
ancilla and prove its optimality. Sec. IV is devoted to different
forms of representation of the introduced machine, in particular
cases giving other universal machines discussed in the literature,
in particular, the symmetric quantum entangler
\cite{SymmEnt,SymmEntExper}, the symmetric
\cite{UCM,UCMopt,UCMnetw,UCMd,UCMdopt,UCMexper} and asymmetric
\cite{AsymmClone,AsymmClone2} quantum cloners, the quantum
information distributor \cite{QID} and the universal-NOT (UNOT)
gate \cite{UNOT,UNOTexper}. In Sec. V we turn to the simplest case
of two-level systems (qubits) and describe a quantum circuit
realizing the universal entangling transformation. In the same
section the asymmetric entangler is compared to the symmetric one,
introduced by Bu\u zek and Hillery, and the differences are
discussed.

\section{Definition of machine}

Let us now define precisely the machine we are interested in. We
consider a signal system $S$ having a state space $\mathcal{H}_S$
of $d$ dimensions and initially prepared in an unknown state
$\ket{\psi}$. Another quantum system $A$ with a state space
$\mathcal{H}_A$ of $d_a$ dimensions is called \textit{ancilla}; it
is initially prepared in a definite state $\ket{Blank}$. We define
quantum entangling machine as a physical device which takes as
input these two systems and produces as output two systems $S'$
and $A'$ with the same number of dimensions of state space $d$ and
$d_a$ respectively. In the future we will often omit primes and
call the output systems $S'$ and $A'$ ``signal'' and ``ancilla''
as well, though they are not necessarily the same physical objects
which entered the machine. We demand for our machine that the
joint state of signal and ancilla at the output is a pure state
$\ket{\Psi}_{SA}$, which generally is not a product of two local
states, i.e. a state for $S'$ and a state for $A'$. To obtain a
universal (input state independent) quantum entangling machine we
demand that the degree of entanglement contained in
$\ket{\Psi}_{SA}$ is independent of the input signal state
$\ket{\psi}$. At this step we need to implement a quantitative
measure of bipartite entanglement. Fortunately, for pure state
there is a good measure of entanglement \cite{Bennet96a}, defined
as von Neumann entropy of the signal system alone:
\begin{equation}\label{E}
E=-Tr\left\{\rho_S\log_2\rho_S\right\},
\end{equation}
where $\rho_S$ is the density operator of the signal at the output
of the entangling machine:
\begin{equation}\label{rhos}
\rho_S=Tr_A\left\{\sket{\Psi}{SA}\;\sbra{\Psi}{SA}\right\}.
\end{equation}
The entanglement $E$ defined in Eq. (3) varies within the limits
$0\le E\le \log_2{d}$, with $E=\log_2{d}$ realized for maximally
entangled systems and $E=0$ for statistically independent ones.
Now the \textit{condition of universal entanglement} can be
formulated as follows: a machine should produce the same amount of
entanglement $E$ for any input signal state $\ket{\psi}$. However,
this condition alone can be satisfied by a trivial machine with
$d_a=d$, which discards the input and produces as output the
maximally entangled state of signal and ancilla:
\begin{equation}\label{max}
\ket{\Psi_{max}}_{SA}=\frac1{\sqrt{d}}\sum_{k=0}^{d-1}\ket{\psi_k}_S\ket{\phi_k}_A,
\end{equation}
where $\ket{\psi_k}_S$ and $\ket{\phi_k}_A$ are two sets of
orthonormal vectors in the state spaces of signal and ancilla
respectively. The entanglement of the output is $E=\log_2{d}$ and
of course it does not depend on the input state which is simply
discarded. The same argument can be applied to the antisymmetric
entangler of Ref. \cite{Alber} (see discussion in Ref.
\cite{SymmEnt}). To obtain a non-trivial machine it is necessary
that the output state, but not the degree of its entanglement, be
related in some way to the unknown input state of the signal. The
most simple and natural way to do so is to demand some
``similarity'' between the output and the input of the signal
system: the signal output should in some sense ``resemble'' the
input, so that the latter cannot be discarded. A natural measure
of similarity of two quantum states is the so-called fidelity
\cite{Jozsa94}. In our case the fidelity between the signal output
and the signal input is defined as
\begin{equation}\label{F}
F=\bra{\psi}\rho_S\ket{\psi},
\end{equation}
and we may formulate \textit{the condition of universal
similarity} as follows: a machine should change the signal system
in such way that the fidelity of the signal output with respect to
the signal input is constant for any input signal state
$\ket{\psi}$.

Thus we define an AUEM as a physical device having as input
systems $S$ and $A$ in states $\ket{\psi}$ and $\ket{Blank}$
respectively and producing as output systems $S'$ and $A'$ in a
pure state $\ket{\Psi}_{SA}$ such that both entanglement Eq.
(\ref{E}) and fidelity Eq. (\ref{F}) are independent of the state
$\ket{\psi}$. We suppose that the entanglement is non-zero and the
fidelity satisfies $\frac1d<F<1$. The fidelity equal to or below
$\frac1d$ is not interesting because in this case the output is
not more similar to the input than the totally mixed state with
the density matrix $\mathbb{I}/d$, where $\mathbb{I}$ is the unity
matrix. We expect that $E(F)$ will be a decreasing function of
$F$, i.e. that we need to ``pay'' for the entanglement by a
degradation of the signal state fidelity. The obvious limiting
points are $E(\frac1d)=\log_2{d}$ (trivial machine with the output
given by Eq.~(\ref{max})) and $E(1)=0$ (no interaction).

The most important question now is if the machine defined in this
way is possible to realize by physical means. There is a number of
quantum machines which are defined by general demands imposed on
the outputs and which proved to be ``impossible'', i.e. not
allowed by the laws of quantum mechanics, for example the perfect
cloning machine \cite{PerfClone} and the perfect universal-NOT
gate \cite{UNOT}. A number of other machines proved to be
``possible'', e.g. the universal (imperfect) cloning machine
\cite{UCM}, the asymmetric cloning machine \cite{AsymmClone},
phase-covariant cloners \cite{CovarClone}, and the (imperfect)
universal-NOT gate \cite{UNOT}. If AUEM would prove to be
possible, another important problem would be to find the optimal
machine, producing maximal degree of entanglement for a given
degradation of the signal state fidelity.

\section{Properties of optimal machine}

Let us suppose that the entangling machine defined in the previous
section is possible and analyze the properties of the optimal one.
While deducing the properties of the optimal machine we will find
its explicit form and thus prove its existence.

\subsection{Purity}

In the definition of the entangling machine we demanded that the
output state of signal and ancilla is pure. Let us now show that
if this state is mixed, entanglement is not greater. Suppose that
the joint state of the signal and the ancilla at the output of the
entangling machine is a mixed state $\rho_{SA}$. The entanglement
of this state can be measured by such a measure as the
entanglement of formation \cite{Bennet96b}, which is found in the
following way. First we unravel the state $\rho_{SA}$, i.e.
represent it as a sum of pure states with some weights, then
calculate the entanglement of each pure state according to
Eq.~(\ref{E}) and take a weighted sum of them, which will give us
the entanglement of that unravelling. The minimum of this quantity
over all possible unravelling of a mixed state is called its
entanglement of formation. For pure states it coincides with the
entanglement given by Eq.~(\ref{E}). It has been proven that the
entanglement of formation is the upper bound for all other
measures of entanglement \cite{Horodecki}.

Let us denote the entanglement of formation of $\rho_{SA}$ as
$E'$. Now we wish to show that for the same output signal state
$\rho_{S}=Tr_A{\rho_{SA}}$ we can construct a machine producing
not less entanglement than $E'$. The output state $\rho_{SA}$, as
any mixed state, can be purified on a larger state space: a second
ancilla $Z$ can be added and a pure state $\ket{\Lambda}_{SAZ}$
can be found on the state space of the signal and both ancillas,
such that
$\rho_{SA}=Tr_{Z}\left\{\ket{\Lambda}\bra{\Lambda}\right\}$. It
can be easily proven from the convexity of von Neumann entropy
that $E'$ is not greater than the entanglement between $S$ and
$AZ$ in $\ket{\Lambda}_{SAZ}$ (see the second Lemma in
Ref.~\cite{Bennet96b}). It means that for any mixed output state
$\rho_{SA}$ we always can construct another entangling machine
with a bigger ancilla and pure output $\ket{\Lambda}_{SAZ}$, which
produces not less entanglement and results in the same
transformation of the signal state.

\subsection{Output state of the optimal machine}

Now we analyze the structure of the pure output state of our
machine. As any bipartite pure state, it can be written in the
form of Schmidt decomposition
\begin{equation}
\ket{\Psi}_{SA}=\sum_{k=1}^d{\lambda_k\ket{\psi_k}_S\ket{\phi_k}_A},
\end{equation}
where $\{\ket{\psi_k}_S,k=1,...,d\}$ is an orthonormal basis in
$\mathcal{H}_S$, $\{\ket{\phi_k}_A,k=1,...,d\}$ is an orthonormal
set of vectors in $\mathcal{H}_A$, whose dimensionality is $d_a\ge
d$, and $\lambda_k$ are some complex numbers (Schmidt
coefficients), satisfying the normalization condition
\begin{equation}\label{lambda}
\sum_{k=1}^d{|\lambda_k|^2}=1.
\end{equation}
We accept that the vectors are numbered so that
$|\lambda_1|\ge|\lambda_2|\ge...\ge|\lambda_d|$. The input signal
state can be decomposed in the basis
$\{\ket{\psi_k}_S,k=1,...,d\}$ as:
\begin{equation}
\ket{\psi}=\sum_{k=1}^d{c_k\ket{\psi_k}},
\end{equation}
where $c_k$ are some complex coefficients, satisfying the
normalization condition
\begin{equation}\label{c}
\sum_{k=1}^d{|c_k|^2}=1.
\end{equation}

Now the fidelity, Eq. (\ref{E}), and the entanglement, Eq.
(\ref{F}), can be expressed as
\begin{equation}\label{Fsum}
F=\sum_{k=1}^d{|\lambda_k|^2|c_k|^2},
\end{equation}
\begin{equation}
E=H(|\lambda_1|^2,|\lambda_2|^2,...,|\lambda_k|^2),
\end{equation}
where
\begin{equation}
H(x_1,x_2,...,x_d)=-\sum_{k=1}^d{x_k\log_2{x_k}},
\end{equation}
is Boltzmann's H-function. Now our task is to find such form of
the output state which (i) maximizes $E$ for given $F$ and (ii)
maximizes $F$ for given $E$.

We start with solving the first problem. Let us consider $F$ and
$E$ as functions on the coordinate space $\mathcal{V}$ created by
$\{|\lambda_k|^2\}$ and $\{|c_k|^2\}$, satisfying the
normalization conditions Eqs. (\ref{lambda}), (\ref{c}) and the
ordering of $\lambda$'s. Let us define on $\mathcal{V}$ a subspace
$\mathcal{V}_0$ by $|c_1|^2=1$. Now, for given $F$ on
$\mathcal{V}_0$: $|\lambda_1|^2=F$ and the entanglement is
\begin{equation}
E_0=-F\log_2{F}+H(|\lambda_2|^2,|\lambda_3|^2,...,|\lambda_d|^2).
\end{equation}
The H-function is maximal when all its arguments are equal. Let us
denote by $\mathcal{V}_1$ the one-parameter subspace of
$\mathcal{V}_0$ with
$|\lambda_2|^2=|\lambda_3|^2=...=|\lambda_d|^2=\left(1-|\lambda_1|^2\right)/(d-1)$.
On this subspace $E_0$ reaches its maximum $E_0^{max}=h_d(F)$,
where the function $h_d(F)$ is defined as
\begin{equation}\label{hprime}
h_d(F)=-F\log_2{F}-(1-F)\log_2{\frac{1-F}{d-1}},
\end{equation}
and has the meaning of maximum of the H-function over $d$
arguments summing up to unity, with the fixed maximal argument
$F$. This function is strictly decreasing for $1/d<F<1$, because
its derivative $h'_d(F)=\log_2{[(1-F)/(Fd-F)]}$ is strictly
negative in this region. On the entire space $\mathcal{V}$ the
relation $|\lambda_1|^2\ge F$ holds (from Eq. (\ref{Fsum})) and
for any given $\lambda_1$, the entanglement is bounded above by
the value $h_d(|\lambda_1|^2)$ (from the meaning of the latter).
Since $h_d(F)$ is a decreasing function of $F$, it follows that
$E\le h_d(|\lambda_1|^2)\le h_d(F)=E_0^{max}$, i.e. $E_0^{max}$ is
the entanglement maximum for the given fidelity.

Now we turn to the second problem. Let us fix the entanglement
$E_1$ satisfying $0<E_1<\log_2{d}$. We define the fidelity $F_1$
by equation $h_d(F_1)=E_1$, which has a unique solution, since
$h_d(F)$ is strictly decreasing on $1/d<F<1$. This fidelity
corresponds to a point in the subspace $\mathcal{V}_1$ (see
above). Let us prove that $F_1$ is the fidelity maximum for the
given entanglement. Suppose that there is a point on
$\mathcal{V}$, giving the fidelity $F_2>F_1$ and entanglement
$E_2$. The maximal value of $E_2$ is given by $h_d(F_2)$, as
proven above. Since the function $h_d(F)$ is \textit{strictly}
decreasing on $1/d<F<1$, it follows that $max(E_2)<E_1$, that is,
the value $E_1$ is unreachable for fidelity higher than $F_1$.
This completes the proof.

Summing up, we see that there is a one-parameter subspace
($\mathcal{V}_1$), where both conditions (i) and (ii) are
satisfied simultaneously. On this subspace
$\ket{\psi}=\ket{\psi_1}$ and the output state can be written as
\begin{equation}\label{out}
\ket{\Psi}_{SA}=\sqrt{F}\ket{\psi}_S\ket{\phi_1}_A+\sqrt{\frac{1-F}{d-1}}\sum_{k=2}^d{\ket{\psi_k}_S\ket{\phi_k}_A},
\end{equation}
where the phases of coefficients are absorbed by
$\{\ket{\phi_k}\}$. Both the fidelity $F$ and the entanglement
$E=h_d(F)$ of the state Eq. (\ref{out}) are independent of the
input state $\ket{\psi}$, and therefore, a machine transforming
$\ket{\psi}_S\ket{Blank}_A$ into the state Eq. (\ref{out}) is an
AUEM and an optimal AUEM. We still need to prove that such a
machine exist. It will be done in the next subsections by deducing
the explicit form of the necessary unitary transformation.

\subsection{Transformation of signal state}

Now we look how the optimal AUEM is ``seen'' by the signal system
alone. For the signal the AUEM acts as a quantum channel, which
transforms its state from a pure state $\ket{\psi}\bra{\psi}$ to a
mixed state $\rho_S$. This transformation can be found by
substituting Eq. (\ref{out}) into Eq. (\ref{rhos}), which gives
\begin{equation}\label{depol}
\rho_S=(1-\pi_s)\ket{\psi}\bra{\psi}+
{\frac{\pi_s}d}\,{\mathbb{I}},
\end{equation}
where $\pi_s$ is connected to $F$ by the relation
$F=1-\pi_s+\pi_s/d$. The quantum channel defined by Eq.
(\ref{depol}) is called ``depolarizing channel'' and the parameter
$\pi_s$ varying from 0 to 1 is known as ``depolarized fraction''.
The depolarizing channel is typically taken as the starting point
for the discussion of universal quantum machines, e.g. quantum
cloners \cite{UCMd,AsymmClone,QID}. In our approach it has been
shown that this channel is optimal for our purpose.

\subsection{Generalized Bell states and generalized Pauli operators}

In future it will be very useful to implement the formalism of
generalized Bell basis and Pauli operators to systems with number
of dimensions greater than 2 \cite{AsymmClone,Fivel}. Let us
consider two quantum systems $X$ and $Y$ having state spaces
$\mathcal{H}_X$ and $\mathcal{H}_Y$ respectively of $d$ dimensions
each. We denote by $\ket{j}_X$ and $\ket{j}_Y$, $j=0,...,d-1$, the
orthonormal bases in $\mathcal{H}_X$ and $\mathcal{H}_Y$
respectively. We introduce a set of $d^2$ generalized Bell (GB)
states on $\mathcal{H}_X\otimes\mathcal{H}_Y$:
\begin{equation}\label{gb}
\ket{\psi_{mn}}_{XY}=\frac1{\sqrt{d}}\sum_{j=0}^{d-1}{e^{2\pi
i(jn/d)}\ket{j}_X\ket{j+m}_Y},
\end{equation}
where the indices $m$ and $n$ take values from 0 to $d-1$. Note
that here and below the summation in all indices is taken modulo
$d$. The states Eq. (\ref{gb}) are normalized to unity and
mutually orthogonal, thus creating an orthonormal basis on
$\mathcal{H}_X\otimes\mathcal{H}_Y$. For $d=2$ these states
coincide with the usual Bell basis of two qubits:
\begin{eqnarray}
\ket{\psi_{00}}=\frac1{\sqrt{2}}\left(\ket{0}\ket{0}+\ket{1}\ket{1}\right)\equiv\ket{\Phi^+},\label{Bell1}\\
\ket{\psi_{01}}=\frac1{\sqrt{2}}\left(\ket{0}\ket{0}-\ket{1}\ket{1}\right)\equiv\ket{\Phi^-},\label{Bell2}\\
\ket{\psi_{10}}=\frac1{\sqrt{2}}\left(\ket{0}\ket{1}+\ket{1}\ket{0}\right)\equiv\ket{\Psi^+},\label{Bell3}\\
\ket{\psi_{11}}=\frac1{\sqrt{2}}\left(\ket{0}\ket{1}-\ket{1}\ket{0}\right)\equiv\ket{\Psi^-}.\label{Bell4}
\end{eqnarray}

We also introduce on each space $\mathcal{H}_X$ and
$\mathcal{H}_Y$ a set of $d^2$ unitary operators, which we call
generalized Pauli (GP) operators:
\begin{equation}\label{u}
U_{m,n}=\sum_{k=0}^{d-1}{e^{2\pi i(kn/d)}\ket{k+m}\bra{k}},
\end{equation}
where $m$ and $n$ again run from 0 to $d-1$. Note that
$U_{00}=\mathbb{I}$. The operators Eq. (\ref{u}) satisfy the
relation
\begin{equation}
U_{m,n}U_{k,l}=U_{m+k,n+l}e^{2\pi i(nk/d)}
\end{equation}
and therefore do not form a group, but it can be shown that they
are related to the so-called Heisenberg group \cite{Fivel}. For
$d=2$, GP operators are proportional to Pauli spin operators:
\begin{eqnarray}
U_{0,1}=\ket{0}\bra{0}-\ket{1}\bra{1}\equiv\sigma_z,\\
U_{1,0}=\ket{1}\bra{0}+\ket{0}\bra{1}\equiv\sigma_x,\\
U_{1,1}=\ket{1}\bra{0}-\ket{0}\bra{1}\equiv-i\sigma_y.
\end{eqnarray}
GP operators are connected to GB states by the following
relations:
\begin{eqnarray}
\left(\mathbb{I}\otimes
U_{m,n}\right)\ket{\psi_{00}}&=&\ket{\psi_{mn}},\\
e^{-2\pi i(mn/d)}\left(U_{-m,n}\otimes
\mathbb{I}\right)\ket{\psi_{00}}&=&\ket{\psi_{mn}},\label{local}
\end{eqnarray}
that is any of $d^2$ GB states can be obtained from
$\ket{\psi_{00}}$ by applying one of $d^2$ GP operators (and
possibly a phase shift) locally to system $X$ or $Y$ alone. This
important property will be used in the future. Another useful
property of GP operators is:
\begin{equation}\label{sumu}
\frac1d\sum_{m,n}{U_{m,n}\rho
U_{m,n}^{\dagger}}=\mathbb{I}Tr\{\rho\},
\end{equation}
which can be deduced from Eq. (\ref{u}) using the identity
\begin{equation}\label{id}
\sum_{n=0}^{d-1}{e^{2\pi i\left[(j-k)n/d\right]}}=d\delta_{jk}.
\end{equation}

Yet another property of GP operators is:
\begin{equation}\label{uu}
\left(U_{m,n}\otimes U_{m,-n}\right)\ket{\psi_{kl}}=e^{-2\pi
i(nk+ml)/d}\ket{\psi_{kl}},
\end{equation}
that is, GB states are eigenstates of operators $U_{m,n}\otimes
U_{m,-n}$ which describe simultaneous local transformations of
systems $X$ and $Y$. From Eqs. (\ref{uu}), (\ref{id}) we get
\begin{equation}
\sum_{m,n}{\left(U_{m,n}\otimes
U_{m,-n}\right)}\ket{\psi_{kl}}=d^2\delta_{k0}\delta_{l0}\ket{\psi_{00}},
\end{equation}
or, taking into account that GB states create a basis on
$\mathcal{H}_X\otimes\mathcal{H}_Y$:
\begin{equation}\label{sumuu}
\sum_{m,n}{\left(U_{m,n}\otimes
U_{m,-n}\right)}=d^2\ket{\psi_{00}}\bra{\psi_{00}}.
\end{equation}

\subsection{Kraus representation}

Using Eq. (\ref{sumu}) we can rewrite the output state of the
depolarizing channel Eq. (\ref{depol}) in the so-called Kraus form
\cite{Preskill}:
\begin{equation}\label{rhoskraus}
\rho_S=\sum_{m=0}^{d-1}{\sum_{n=0}^{d-1}{K_{m,n}\ket{\psi}\bra{\psi}K_{m,n}^\dagger}},
\end{equation}
where
\begin{equation}\label{k}
K_{m,n}=\left\{
\begin{array}{lcl}
aU_{0,0},&\quad &m,n=0,0,\\
bU_{m,n},&\quad &m,n\ne0,0,\\
\end{array}\right.
\end{equation}
are Kraus operators and the parameters $a$ and $b$ determine their
relative weights and are connected to the depolarizing fraction by
the relations
\begin{eqnarray}
a&=&\sqrt{1-\pi_s+\pi_s/d^2},\label{a}\\
b&=&\sqrt{\pi_s}/d.\label{b}
\end{eqnarray}

Kraus representation helps find the form of the output state of
signal and ancilla, namely:
\begin{equation}\label{outkraus}
\ket{\Psi}_{SA}=\sum_{m,n}{\left[\vphantom{\tilde
K}K_{m,n}\ket{\psi}\right]}\ket{\phi_{mn}},
\end{equation}
where $\left\{\ket{\phi_{mn}}\right\}$ is any set of $d^2$
orthonormal vectors in the state space of ancilla. It is
straightforward to verify that substituting Eq. (\ref{outkraus})
in Eq. (\ref{rhos}) gives  Eq. (\ref{rhoskraus}). We see that we
need only $d^2$ dimensions of ancilla to store the possible
transformations of the signal state. Thus we can use for our
ancilla a pair of $d$-level systems $X$ and $Y$ and identify the
states $\left\{\ket{\phi_{m,n}}\right\}$ as phase-shifted GB
states on $\mathcal{H}_A\equiv\mathcal{H}_X\otimes\mathcal{H}_Y$:
\begin{equation}\label{phipsi}
\ket{\phi_{mn}}=\left\{
\begin{array}{lcl}
e^{i\varphi}\ket{\psi_{00}},&\quad &m,n=0,0,\\
e^{2\pi i(mn/d)}\ket{\psi_{-m\,-n}},&\quad &m,n\ne0,0,\\
\end{array}\right.
\end{equation}
where $\varphi$ is some phase which will be used for the ``fine
tuning'' of the overall system-ancilla transformation.

Substituting Eqs. (\ref{phipsi}), (\ref{k}) into Eq.
(\ref{outkraus}), using Eq. (\ref{local}) and having in mind that
$U_{0,0}=\mathbb{I}$, we obtain
\begin{eqnarray}
\ket{\Psi}_{SXY}
=\left(ae^{i\varphi}+b\sum_{m,n\ne0,0}{U_{m,n}^{(S)}\otimes
U_{m,-n}^{(X)}}\right)\nonumber\\
\times\ket{\psi}_S\ket{\psi_{00}}_{XY},
\end{eqnarray}
where the upper index shows in which space an operator acts and
for simplicity the identity operators are omitted. Applying Eq.
(\ref{sumuu}) to the last equation we get finally
\begin{equation}\label{form1}
\ket{\Psi}_{SXY}=M\ket{\psi}_S\ket{\psi_{00}}_{XY},
\end{equation}
where the operator $M$ is defined as
\begin{equation}\label{m}
M=\alpha+\beta d\sket{\psi_{00}}{SX}\sbra{\psi_{00}}{SX},
\end{equation}
where the complex parameter $\alpha$ and the real parameter
$\beta$ are defined as
\begin{eqnarray}
\alpha&=&ae^{i\varphi}-b,\label{alpha}\\
\beta&=&bd=\sqrt{\pi_s}\;,\label{beta}
\end{eqnarray}
and satisfy the relation
\begin{equation}\label{alphabeta}
|\alpha|^2+\frac2dRe\{\alpha\}\beta+\beta^2=1,
\end{equation}
following from Eqs. (\ref{a}), (\ref{b}), (\ref{alpha}),
(\ref{beta}). The meaning of the parameter $\beta$ is clear, it is
the square root of the depolarizing fraction of the signal
channel. The meaning of the parameter $\alpha$ will be clarified
below.

\subsection{Existence of AUEM}

Eq. (\ref{form1}) allows us to find the explicit form of AUEM and
thus prove its existence. At first glance, this equation shows us
that we can use the state $\ket{\psi_{00}}_{XY}$ as the initial
state of ancilla $\ket{Blank}$ and apply the transformation $M$ to
obtain the necessary output state. Unfortunately, the operator
$M$, as defined by Eq. (\ref{m}) is generally not unitary on
$\mathcal{H}=\mathcal{H}_S\otimes\mathcal{H}_X\otimes\mathcal{H}_Y$
and therefore does not correspond to a physical process. However,
we could construct a unitary operator on $\mathcal{H}$ suitable
for our needs if $M$ would be unitary on a subspace $\mathcal{W}$
spanned by possible input states of the machine, namely the states
of the form $\ket{k}_S\ket{\psi_{00}}_{XY}$, where
$\left\{\ket{k}_S,k=0,...,d-1\right\}$ is some basis in
$\mathcal{H}_S$. Let us verify this fact. From Eqs. (\ref{form1}),
(\ref{m}) we find:
\begin{equation}\label{form2}
M\ket{k}_S\ket{\psi_{00}}_{XY}=\alpha\ket{k}_S\ket{\psi_{00}}_{XY}+\beta\ket{\psi_{00}}_{SX}\ket{k}_Y.
\end{equation}
Let us denote the subspace spanned by vectors given by Eq.
(\ref{form2}) as $\mathcal{W}'$. We can easily calculate with the
help of Eq. (\ref{alphabeta})
\begin{equation}
\sbra{\psi_{00}}{XY}\sbra{k}SM^{\dagger}M\sket{l}S\sket{\psi_{00}}{XY}=\delta_{k
l},
\end{equation}
that is $M$ is unitary on $\mathcal{W}$ as required. It means that
the orthonormal set of $d$ states on $\mathcal{W}$ is transformed
into an orthonormal set of $d$ states on $\mathcal{W}'$ . We can
define a transformation $U_M$:
$\mathcal{H}\longrightarrow\mathcal{H}$ so that it acts as $M$ on
$\mathcal{W}$ and transforms an orthonormal set of $d^3-d$ states
on $\mathcal{H}\setminus\mathcal{W}$ into an orthonormal set on
$\mathcal{H}\setminus\mathcal{W}'$. This transformation maps one
basis on $\mathcal{H}$ onto another and therefore is unitary and
realizable by physical means.

Thus we have proven the existence of AUEM and have found, that the
optimal machine is realized by restricting our ancilla to a pair
of $d$-level systems $X$ and $Y$, preparing its input state in the
GB state $\ket{\psi_{00}}_{XY}$ and applying to the signal and
ancilla the unitary transformation $U_M$, constructed as described
above.

\subsection{Uniqueness}

How unique is the described optimal machine? Other forms of the
optimal AUEM can be connected only with other Kraus
representations of the depolarizing channel and different ways of
identification of the states $\ket{\phi_{mn}}$ in the state space
of ancilla. It is known that different sets of Kraus operators are
connected by a unitary transformation \cite{Preskill}, that is, if
we find operators $\tilde K_{\mu,\nu}$ such that
\begin{equation}\label{rhoskrausalt}
\rho_S=\sum_{\mu=0}^{d-1}{\sum_{\nu=0}^{d-1}{\tilde
K_{\mu,\nu}\ket{\psi}\bra{\psi}\tilde K_{\mu,\nu}^\dagger}},
\end{equation}
then
\begin{equation}\label{kvk}
\tilde
K_{\mu,\nu}=\sum_{m=0}^{d-1}{\sum_{n=0}^{d-1}{\Lambda_{\mu\nu m
n}K_{m,n}}}
\end{equation}
where $\Lambda_{\mu\nu m n}$ are elements of a unitary matrix
(pairs $(\mu\nu)$ and $(m n)$ being considered as two indexes).
Eq. (\ref{rhoskrausalt}) leads us to the output state
\begin{equation}\label{outkrausalt}
\ket{\Psi}_{SA}=\sum_{\mu,\nu}{\left[\tilde
K_{\mu,\nu}\ket{\psi}\right]\ket{\vartheta_{\mu\nu}}},
\end{equation}
where $\ket{\vartheta_{\mu\nu}}$ is an orthonormal basis on
$\mathcal{H}_X\otimes\mathcal{H}_Y$. Substituting Eq. (\ref{kvk})
into Eq. (\ref{outkrausalt}) and comparing the result with Eq.
(\ref{outkraus}), we find
\begin{equation}
\ket{\phi_{mn}}=\sum_{\mu=0}^{d-1}{\sum_{\nu=0}^{d-1}{\Lambda_{\mu\nu
m n}\ket{\vartheta_{\mu\nu}}}},
\end{equation}
that is, any other form of Kraus representation of the
depolarizing channel can be reduced to that described in the
previous subsection by a unitary transformation of ancilla state
space. Different ways of identification of states
$\ket{\phi_{mn}}$, which form an orthonormal basis on
$\mathcal{H}_A$, are also connected to one another by a unitary
transformation on this space. Note that different values of
$\varphi$ also can be obtained by unitary transformations of
ancilla. Thus, up to a unitary transformation on
$\mathcal{H}_X\otimes\mathcal{H}_Y$, there is only one optimal
AUEM for any given fidelity $F$.

\section{Representations of AUEM}

In the previous section we have solved a \textit{mathematical}
problem of existence and uniqueness of AUEM, namely we have found
the form of optimal AUEM unique up to a unitary transformation of
the ancilla state space. The transformation of
$\mathcal{H}_X\otimes\mathcal{H}_Y$ may mix the states of the two
systems $X$ and $Y$ composing the ancilla and change the form of
the interaction between the signal system and the ancilla. When
looking for a \textit{physical} realization of the machine of our
interest, we may find some forms of states and interactions more
``handy'' than others, so it seems to be very interesting to
investigate different representations of the optimal AUEM
corresponding to different choices of the ancilla transformation.

\subsection{Standard representation}

First we summarize the results of the previous section and
describe the representation of the optimal AUEM which will serve
as the starting point for considering the transformations of the
ancilla state space.

At the input of the machine we have three $d$-level systems: $S$,
$X$ and $Y$. For each system we define the ``standard'' basis
$B_i$: $\left\{\ket{k}_i, k=0,...,d-1\right\}$, where $i=S,X,Y$.
The signal is initially in an unknown state $\ket{\psi}_S$. We
prepare the systems $X$ and $Y$ in a maximally entangled state,
which is the $\ket{\psi_{00}}_{XY}$ state in the basis $B_XB_Y$.
We perform a unitary transformation $U_M$ of all three systems,
which acts on the input state as the operator $M$ defined by Eq.
(\ref{m}) in the basis $B_SB_X$. As a result we obtain the
transformation, which in the basis $B_SB_XB_Y$ is written as
\begin{equation}\label{standard}
\ket{\psi}_S\ket{\psi_{00}}_{XY}\longrightarrow
\alpha\ket{\psi}_S\ket{\psi_{00}}_{XY}
+\beta\ket{\psi_{00}}_{SX}\ket{\psi}_Y,
\end{equation}
with complex $\alpha$ and real $\beta$ satisfying Eq.
(\ref{alphabeta}).

\subsection{Covariance}

Let us suppose that instead of basis $B_S$ we wish to use another
basis in $\mathcal{H}_S$, say $B'_S$: $\left\{V\ket{k}_S,
k=0,...,d-1\right\}$, where $V$ is some unitary transformation on
$\mathcal{H}_S$. This situation can be met in quantum cryptography
\cite{BB84,GisinRev}, where the eavesdropper may use the AUEM for
intercepting the secret message (signal). In the majority of
quantum cryptographic protocols, e.g. BB84 \cite{BB84}, after the
transmission of quantum data a set of vectors is announced from
which the transmitted state $\ket{\psi}_S$ has been chosen. When
calculating the amount of intercepted information it is natural to
use the basis in $\mathcal{H}_S$ comprising this set. The question
arises if we can choose the corresponding bases in $\mathcal{H}_X$
and $\mathcal{H}_Y$ so that the form of transformation given by
Eq. (\ref{standard}) remain unchanged? To find the answer we note
the remarkable property of the $\ket{\psi_{00}}_{SX}$ state,
namely its invariance under the simultaneous change of bases of
systems $S$ and $X$ to bases $B'_S$:
$\left\{\ket{\zeta_k}_S=V\ket{k}, k=0,...,d-1\right\}$ and $B'_X$:
$\left\{\ket{\zeta_k}_X=V^*\ket{k}, k=0,...,d-1\right\}$ where $V$
is a unitary operator and $V^*$ is its complex conjugation in the
basis $B_X$. Indeed, expressing $\ket{\psi_{00}}_{SX}$ in the new
basis we find
\begin{eqnarray}\label{vconjug}
\ket{\psi_{00}}_{SX}
&=&\sum_{m,n}{\sket{\zeta_m}S\sket{\zeta_n}X\sbra{\zeta_n}X\sbra{\zeta_m}S\cdot\ket{\psi_{00}}_{SX}}\nonumber\\
&=&\frac1{\sqrt{d}}\sum_{k,m,n}{v^*_{km}v_{kn}\ket{\zeta_m}_S\ket{\zeta_n}_X}\nonumber\\
&=&\frac1{\sqrt{d}}\sum_{m}{\ket{\zeta_m}_S\ket{\zeta_m}_X},
\end{eqnarray}
where $v_{kn}=\bra{k}V\ket{n}$ are matrix elements of the operator
$V$ in the $B_S$ ($B_X$) basis, satisfying the unitarity relation
\begin{equation}
\sum_{k}{v^*_{km}v_{kn}}=\delta_{mn}.
\end{equation}
It follows from Eq. (\ref{vconjug}) that if we apply
transformation $V$ to the bases $B_S$ and $B_Y$ and transformation
$V^*$ to the basis $B_X$, then in the new bases the states
$\ket{\psi_{00}}_{SX}$ and $\ket{\psi_{00}}_{XY}$ will preserve
their form, and the entire transformation Eq. (\ref{standard})
will remain invariant.

Alternatively, we can express this property as covariance of the
output state $\ket{\Psi}_{SXY}$ with the input state
$\ket{\psi}_S$, namely if we have the state $U\ket{\psi}_S$ at the
signal input, we get the state $U\otimes U^*\otimes
U\ket{\Psi}_{SXY}$ at the output, where $U$ is any unitary
transformation on a $d$-level space and the asterisk again stands
for complex conjugation in the $B_X$ basis. Such a covariance
follows from the invariance of the state $\ket{\psi_{0,0}}_{SX}$
under the transformation $U\otimes U^*$:
\begin{equation}\label{uconjug}
U\otimes U^*\ket{\psi_{00}}_{SX}=\ket{\psi_{00}}_{SX},
\end{equation}
which can be proven in the same way as Eq. (\ref{vconjug}).

\subsection{The local output states}
\label{conj}

Now we calculate from Eq. (\ref{standard}) the output states of
the three systems under consideration by tracing out the other
two:
\begin{equation}
\rho_S=\left[|\alpha|^2+\frac2d
Re\{\alpha\}\beta\right]\sket{\psi}S\sbra{\psi}S+\beta^2\frac{\mathbb{I}}d,
\end{equation}
\begin{equation}
\rho_X=\frac2d
Re\{\alpha\}\beta\sket{\psi^*}X\sbra{\psi^*}X+\left[|\alpha|^2+\beta^2\right]\frac{\mathbb{I}}d,
\end{equation}
\begin{equation}
\rho_Y=\left[\frac2d
Re\{\alpha\}\beta+\beta^2\right]\sket{\psi}Y\sbra{\psi}Y+|\alpha|^2\frac{\mathbb{I}}d.
\end{equation}
We see that the output of system $Y$, like the signal output,
represents a depolarizing channel with respect to the signal
input, but characterized by the depolarizing fraction
$\pi_y=|\alpha|^2$ instead of $\pi_s=\beta^2$. The output of
system $Y$ can be considered as an inaccurate copy, or ``clone''
of the signal system. The quality of the clone is optimal for some
values of the parameters characterizing the AUEM, which are
discussed below.

We also note that the output of system $X$ is the state
$\ket{\psi^*}\bra{\psi^*}$, emerging from a depolarizing channel
with the depolarized fraction $\pi_x=|\alpha|^2+\beta^2$. This
output realizes also a very important quantum machine - quantum
conjugator. Generally, the map
$\ket{\psi}\longrightarrow\ket{\psi^*}$ cannot be realized by
physical means because it is not unitary. However such a
transformation can be realized approximately as map
$\ket{\psi}\longrightarrow\rho$, with some fidelity
$F_{conj}=\bra{\psi^*}\rho\ket{\psi^*}<1$. Such a map has been
recently analyzed for the quantum continuous variables
\cite{ContConj}. In our case $F_{conj}=|\alpha+\beta|^2/d$; its
optimal value will be found later on. For the case of even $d$ the
problem of conjugation is closely related to the problem of
generating a state orthogonal to a given state, i.e. a map:
$\ket{\psi}\longrightarrow\ket{\psi^\perp}$, where
$\ket{\psi^\perp}$ is a state orthogonal to $\ket{\psi}$. Such a
map is also possible only approximately, but if we could realize
(approximately) the conjugated state $\ket{\psi^*}$ then we could
unitarily transform it into the state
$\ket{\psi^\perp}=\sigma_y^{01}\sigma_y^{23}...\sigma_y^{(d-2)(d-1)}\ket{\psi^*}$,
where the unitary rotation
$\sigma_y^{kl}=-i\ket{k}\bra{l}+i\ket{l}\bra{k}$ is an analog of
Pauli operator $\sigma_y$ for two vectors of the $B_X$ basis. It
is easy to check that the obtained state is orthogonal to
$\ket{\psi}$.

\subsection{Asymmetric cloner representation ($\varphi=0$)}

The minimal value of $\pi_y$ for fixed $F$ (and therefore, $a$ and
$b$) is reached when $\varphi=0$ and $\alpha$ is real. In this
case the optimal AUEM acts as the asymmetric cloner introduced by
Cerf \cite{AsymmClone} and the output of the system $Y$ can be
considered as the best possible ``clone'' of the signal input for
the given value of the signal channel fidelity. With decreasing
$F$ the quality of clone increases. The relation between the
depolarized fractions of the signal and the system $Y$ is given by
Eq.~(\ref{alphabeta}) with real $\alpha$ and is exactly that for
the asymmetric cloner. The quantum information distributor,
suggested in Ref. \cite{QID}, also realizing asymmetric cloning,
provides a transformation defined by Eq. (\ref{standard}) but with
real parameters $\alpha$ and $\beta$.

\subsection{Symmetric cloner and UNOT gate. Symmetric entangler. Optimal conjugator ($\varphi=0$,
$\alpha=\beta$)}\label{se}

Let us analyze the particular case of asymmetric cloner
($\varphi=0$) where both clones ($S'$ and $Y'$) arise with the
same fidelity with respect to the input signal state, i.e. where
$\alpha=\beta=\sqrt{d/(2d+2)}$. In this case the AUEM works as the
symmetric cloner, and the systems $S'$ and $Y'$ give two
``clones'' of the input $S$ with the same fidelity
$F=(d+3)/(2d+2)$, which has been proven to be optimal
\cite{UCMdopt}.

In the case where all systems $S$, $X$, and $Y$ are qubits ($d=2$)
AUEM is closely related to the machine introduced by Bu\u zek and
Hillery \cite{UCM}. It has been introduced first as a symmetric
quantum cloner but it has proved later that it works at the same
time as the optimal UNOT-gate \cite{UNOT} and optimal symmetric
entangler \cite{SymmEnt}. This machine has been recently realized
in an experiment \cite{SymmEntExper}. The only difference between
the AUEM with $\varphi=0$, $\alpha=\beta$ and the symmetric cloner
is that the output $X'$ of AUEM should be rotated by operator
$-i\sigma_y$. In this case the output state of the machine is
obtained from Eq. (\ref{standard}) with $\alpha=\beta=1/\sqrt3$:
\begin{equation}\label{symm1}
\ket{\Psi_{out}}=\frac1{\sqrt6}\left[\ket{\psi}\left(\ket{10}-\ket{01}\right)
+\left(\ket{01}-\ket{10}\right)\ket{\psi}\right],
\end{equation}
where $\ket{01}=\ket{0}\ket{1}$ and the kets are written in the
order $SXY$ in the basis $B_SB_XB_Y$. Now, using the invariance of
the singlet state with respect to the change of basis, we can
write it as
$\ket{10}-\ket{01}=\ket{\psi}\ket{\psi^\perp}-\ket{\psi^\perp}\ket{\psi}$,
where $\ket{\psi^\perp}=i\sigma_y\ket{\psi^*}$ is a state
orthogonal to $\ket{\psi}$, and rewrite Eq. (\ref{symm1}) as
\begin{equation}\label{symm2}
\ket{\Psi_{out}}=-\sqrt{\frac23}\sket{\psi}S\sket{\psi^\perp}X\sket{\psi}Y+
\frac1{\sqrt3}\sket{\{\psi,\psi^\perp\}}{SY}\sket{\psi}X,
\end{equation}
where
$\ket{\{\psi,\psi^\perp\}}=2^{-1/2}(\ket{\psi}\ket{\psi^\perp}+\ket{\psi^\perp}\ket{\psi})$
is the symmetric entangled state of two qubits. It has been shown
that the machine producing the output state Eq. (\ref{symm2}) is
the optimal symmetric entangler of $S$ and $Y$ \cite{SymmEnt}, the
optimal symmetric cloner (outputs $S$ and $Y$) \cite{UCMopt} and
the optimal UNOT gate (output $X$) \cite{UNOT}. The latter has a
fidelity $F_{UNOT}=2/3$, which is equal to the fidelity obtained
by the optimal estimation of $\ket{\psi}$ and preparation of
$\ket{\psi^\perp}$.

We return to the case of arbitrary $d$. It seems that complex
conjugation is a natural generalization of UNOT operation to the
case of $d$-dimensional systems (see discussion in subsection
\ref{conj}). The fidelity of the ``conjugated clone'' $X'$ is
maximal for $\varphi=0$ and $\alpha=\beta$ (the same values as for
the realization of the symmetric cloner), in which case
$F_{conj}=2/(d+1)$ which again is equal to the fidelity of the
``estimation and preparation'' method \cite{UCMd,Esteem1d}. It is
unknown to us if the fidelity of conjugation reached by the AUEM
is that maximal allowed by the laws of Nature.

\subsection{Minimal-interaction representation}

So far we have considered the representations of AUEM coinciding
with other universal machines proposed in the literature. Now we
introduce a new representation having the advantage of minimizing
the interaction between the signal and ancilla. To this end we
look for a possibility to make unitary the operator $M$, defined
by Eq. (\ref{m}). If it would prove to be possible, the entangling
interaction could comprise the systems $S$ and $X$ only.

It is easy to see that in the basis of GB states of systems $S$
and $X$,
$M=diag\{ae^{i\varphi}-b+bd^2,ae^{i\varphi}-b,...,ae^{i\varphi}-b\}$.
For this operator to be unitary it is necessary and enough that
\begin{eqnarray}
|ae^{i\varphi}-b+bd^2|&=&1,\\
|ae^{i\varphi}-b|&=&1.
\end{eqnarray}
Both equations are satisfied at $\varphi=\varphi_0$, where
\begin{equation}\label{phi0}
\cos{\varphi_0}=-\frac{d^2-2}2\sqrt{\frac{1-F}{(d^2-1)F-d+1}},
\end{equation}
which is possible for
\begin{equation}\label{flim}
F\ge1-\frac{4(d-1)}{d^3}.
\end{equation}
For example, for $d=2$ the fidelity of the signal may take any
value above $\frac12$; for $d=3$ the fidelity must be
$\frac{19}{27}$ or higher.

If the fidelity is high enough and the phase $\varphi$ is chosen
according to Eq. (\ref{phi0}), the operator $M$ on
$\mathcal{H}_S\otimes\mathcal{H}_X$ has the unitary form
\begin{equation}
M=e^{i\theta_0}e^{i\theta\ket{\psi_{00}}\bra{\psi_{00}}},
\end{equation}
where
\begin{eqnarray}
\cos{\theta_0}&=&-\sqrt{\frac{d^3(1-F)}{4(d-1)}},\\
\cos{\theta}&=&1-\frac{d^3(1-F)}{2(d-1)}.\label{theta}
\end{eqnarray}
In this case an optimal AUEM can be realized by implementing the
interaction of systems $S$ and $X$ only. This may be very
important in some practical applications of AUEM, e.g. in quantum
cryptography, where AUEM can be used for eavesdropping.

Note, that different values of $\varphi$ can be obtained by
shifting the phase of the state $\sket{\psi_{00}}{XY}$, and
leaving the rest of $\mathcal{H}_X\otimes\mathcal{H}_Y$ intact.
Therefore the unitary transformation realizing AUEM for $F$
satisfying Eq. (\ref{flim}) and arbitrary $\varphi$ can be written
as
\begin{equation}\label{um}
U_M=e^{i\theta_0}e^{i(\varphi-\varphi_0)\ssket{\psi_{00}}{XY}\ssbra{\psi_{00}}{XY}}
e^{i\theta\ssket{\psi_{00}}{SX}\ssbra{\psi_{00}}{SX}}.
\end{equation}
It can be checked directly that this operator transforms the
possible input states in exact accordance with Eq.
(\ref{standard}):
\begin{eqnarray}
U_M\sket{\psi}S\sket{\psi_{00}}{XY}=e^{i(\varphi-\varphi_0)\ssket{\psi_{00}}{XY}\ssbra{\psi_{00}}{XY}}\nonumber\\
\times\left[(ae^{i\varphi_0}-b)\sket{\psi}S\sket{\psi_{00}}{XY}+bd\sket{\psi_{00}}{SX}\sket{\psi}Y\right]\nonumber\\
=(ae^{i\varphi}-b)\sket{\psi}S\sket{\psi_{00}}{XY}+bd\sket{\psi_{00}}{SX}\sket{\psi}Y.
\end{eqnarray}

The minimal-interaction representation, given by Eq. (\ref{um})
shows that an AUEM can be realized by sequential application of
the same physical process first to systems $S$ and $X$ and then to
the systems $X$ and $Y$. Let us denote by $\textbf{G}(\varphi)$ a
quantum gate, acting on two $d$-level systems $A$ and $B$ and
shifting the phase of the GB state $\sket{\psi_{00}}{AB}$ by
$\varphi$, while doing nothing for the other GB states. Then, up
to irrelevant overall phase, the optimal AUEM can be realized by
applying $\textbf{G}(\theta)$ to $S$ and $X$ first and
$\textbf{G}(\varphi-\varphi_0)$ to $X$ and $Y$ afterwards. A
concrete example of such a quantum circuit for $d=2$ will be
discussed in the next section.

\section{Entangling machine for one qubit}

In this section we consider the simplest case of $d=2$, where all
three systems $S$, $X$ and $Y$ are represented by two-level
systems or ``qubits''. We describe the quantum circuit realizing
the optimal AUEM, compare the performance of the AUEM to that of
the symmetric entangler and show how the AUEM can be applied to
eavesdropping on a quantum cryptographic line.

\subsection{Quantum circuit}

\begin{figure}[h]
\includegraphics[scale=1.0]{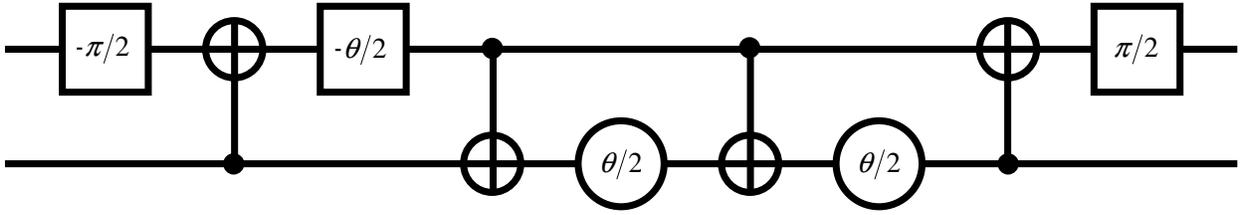}
\caption{\label{g} Quantum circuit realizing the gate
$\textbf{G}(\theta)$. The Bell states of two qubits are its
eigenstates. It shifts the phase of $\Phi^+$ state by $3\theta/4$
and the phases of the other three Bell states by $-\theta/4$.
Two-qubit gates are CNOT gates, while one-qubit gates are
rotations around $z$ (squares) or $y$ (circles) axis by the
specified angle.}
\end{figure}

A quantum circuit realizing the optimal AUEM for one qubit can be
build with the help of the two-qubit circuit depicted in
Fig.~\ref{g}. Horizontal lines represent qubits, vertical lines
are two-qubit CNOT gates, and the squares and circles represent
one-qubit rotations $\exp{(-i\sigma_z\xi/2)}$ and
$\exp{(-i\sigma_y\xi/2)}$ respectively by the specified angle
$\xi$. It is straightforward to verify that the four Bell states
of the input qubits, defined by Eqs. (\ref{Bell1}-\ref{Bell4}) are
transformed as follows:
\begin{eqnarray}
\ket{\Phi^+}&\longrightarrow&e^{i3\theta/4}\ket{\Phi^+},\\
\ket{\Phi^-}&\longrightarrow&e^{-i\theta/4}\ket{\Phi^-},\\
\ket{\Psi^\pm}&\longrightarrow&e^{-i\theta/4}\ket{\Psi^\pm},
\end{eqnarray}
i.e. they are not mixed with one another, each acquiring a phase
shift in such a way that there is a phase difference of $\theta$
between $\ket{\Phi^+}$ and the other three Bell states. It means
that this circuit realizes the gate $\textbf{G}(\theta)$ defined
in the previous Section. This circuit alone realizes an optimal
AUEM, if one of its entries  is considered as input (qubit $S$)
and the other (qubit $X$) is prepared in the Bell state
$\sket{\Phi^+}{XY}$ with the third qubit ($Y$). The fidelity of
the machine is connected to the parameter $\theta$ by Eq.
(\ref{theta}).

\begin{figure}[h]
\includegraphics[scale=1.0]{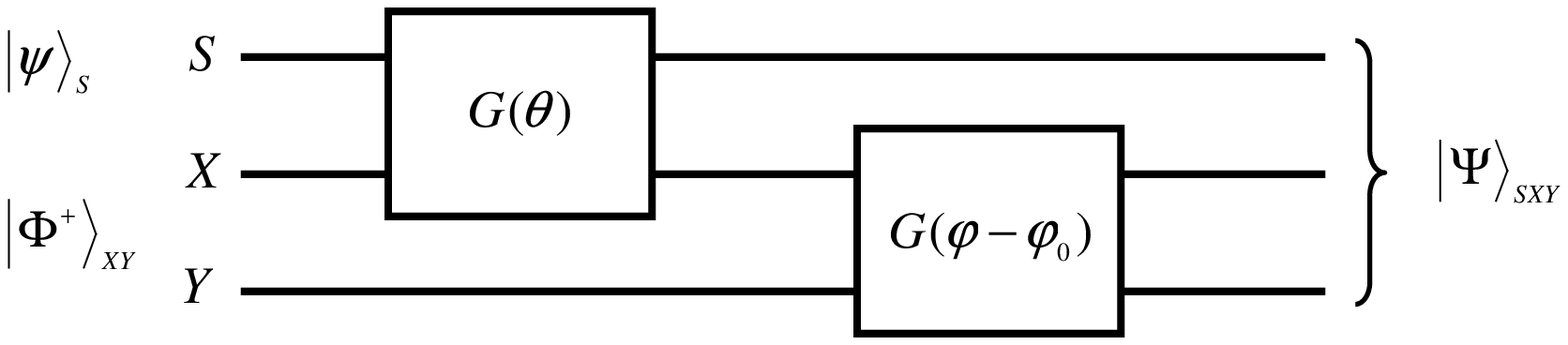}
\caption{\label{auem} AUEM for one qubit. $S$ is the signal qubit
in an unknown state. Qubits $X$ and $Y$ compose the 4-level
ancilla and are initially prepared in the $\Phi^+$ Bell state. The
gate $G(\theta)$ is shown in Fig.~\ref{g}. The angles $\theta$ and
$\varphi_0$ are connected to the fidelity $F$ by Eqs.
(\ref{theta}), (\ref{phi0}) with $d=2$. $\varphi$ is a free
parameter. The output state is the pure entangled state of the
three qubits.}
\end{figure}

Other representations of AUEM for one qubit can be obtained by
concatenating this circuit with a similar one applied to the
qubits $X$ and $Y$ and with $\theta$ replaced by
$\varphi-\varphi_0$ (see Fig.~\ref{auem}). For example, the
asymmetric cloner \cite{AsymmClone} may be realized in this way by
putting $\varphi=0$.

The symmetric entangler, acting at the same time as the symmetric
cloner and the universal-NOT gate, is realized by putting
$\varphi=0$, $\theta=\arccos{(1/3)}$. The circuit depicted in
Fig.~\ref{auem} can be compared to the circuits suggested for the
universal (symmetric) cloning machine \cite{UCMnetw}, where both
ancillary qubits interact with the signal. Our scheme has the
advantage of minimizing the number of qubits involved into the
interaction with the signal system.

\subsection{Comparison with the symmetric entangler}

The performance of the optimal AUEM for one qubit, can be compared
to that of the symmetric entangler of Bu\u zek and Hillery
\cite{SymmEnt}. The symmetric entangler (see subsection~\ref{se})
is a machine having two qubits as input, one of them being the
signal qubit in an unknown state $\ket{\psi}$ and another one
being a specially prepared ancilla. The output of the machine is a
pair of qubits in a mixed entangled state obtained from
Eq.~(\ref{symm2}) by tracing out the system $X$:
\begin{eqnarray}\label{rho}
\rho_{SY}&=&\frac16\left(\ket{\psi}\ket{\psi^\perp}+\ket{\psi^\perp}\ket{\psi}\right)
\left(\bra{\psi}\bra{\psi^\perp}+\bra{\psi^\perp}\bra{\psi}\right)\nonumber\\
&&+\frac23\ket{\psi}\ket{\psi}\bra{\psi}\bra{\psi},
\end{eqnarray}
where the first vector in each pair of kets or bras is related to
the signal qubit and the second one -- to the ancillary qubit.

\begin{figure}[h]
\includegraphics[scale=0.8]{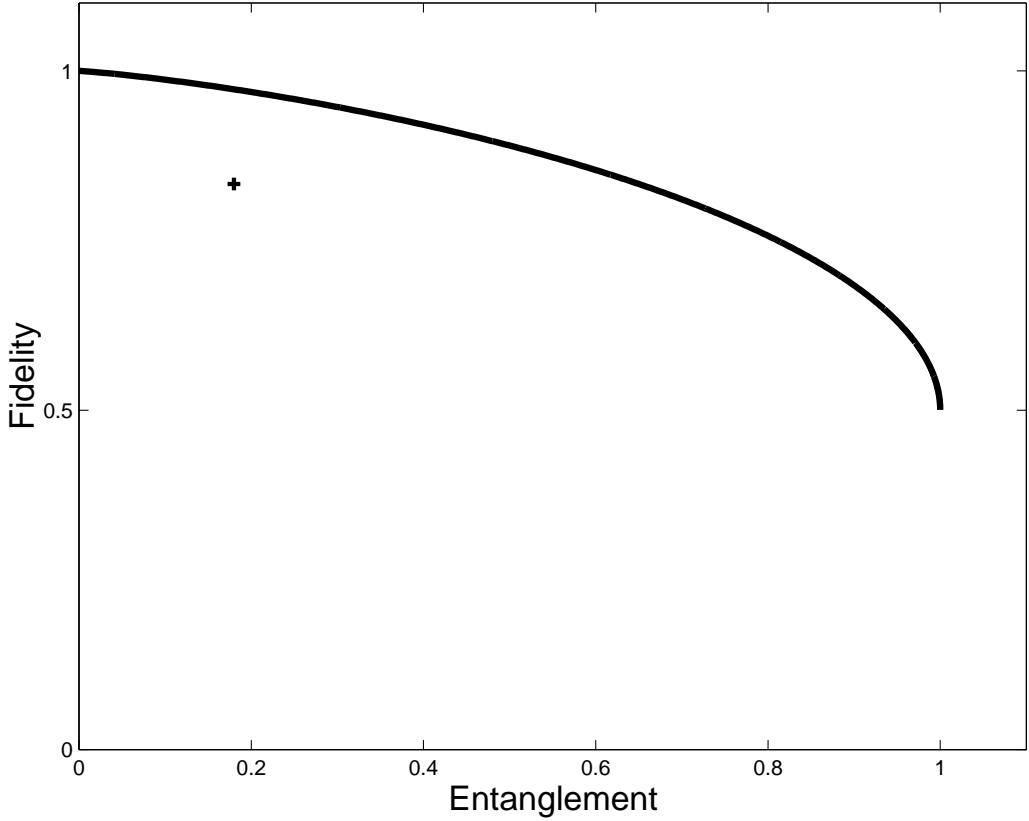}
\caption{\label{FvE}The output-to-input fidelity of the signal
qubit as a function of the entanglement between the signal qubit
and the ancilla, at the output of AUEM (solid line). The symmetric
entangler of Ref.~\cite{SymmEnt} is shown by a cross.}
\end{figure}

Using the Wooters' method \cite{WootersConc} we calculate the
amount of entanglement of formation contained in the output state,
which is the upper bound for all other measures of entanglement.
The concurrence of the state Eq. (\ref{rho}) is $C=\frac13$ and
the entanglement of formation is equal to
$h_2\left(\frac12(1+\sqrt{1-C^2})\right)\approx0.19$ ebits. The
output-to-input fidelity of the signal qubit can be easily found
equal to $\frac56$. These quantities can be compared to the
corresponding quantities of the optimal AUEM. For the price of
disturbance $\frac16$ the AUEM produces $E=h_2(\frac56)=0.65$
ebits of entanglement, i.e. it is several times more effective
than the symmetric entangler. It should be remembered also that
the AUEM, in contrast to the symmetric entangler, allows us to
vary the disturbance of the signal, finding the optimal trade-off
between the fidelity and the entanglement (Fig.~\ref{FvE}), which
is very important for such applications as eavesdropping in
quantum cryptography.

\subsection{Application of AUEM to eavesdropping}

Let us show that the optimal AUEM for one qubit realizes the
interaction necessary for the optimal eavesdropping in the
so-called six state protocol of quantum cryptography. In this
protocol the value of a bit is encoded into the state of a qubit
$S$, chosen from three bases: the ``rectilinear'' one, created by
two orthonormal vectors $\ket{0}$ and $\ket{1}$, the ``diagonal''
one, created by
\begin{eqnarray}
\ket{\bar0}=\frac1{\sqrt{2}}\left(\ket{0}+\ket{1}\right),\\
\ket{\bar1}=\frac1{\sqrt{2}}\left(\ket{0}-\ket{1}\right),
\end{eqnarray}
and the ``circular'' one:
\begin{eqnarray}
\ket{0'}=\frac1{\sqrt{2}}\left(\ket{0}+i\ket{1}\right),\\
\ket{1'}=\frac1{\sqrt{2}}\left(\ket{0}-i\ket{1}\right).
\end{eqnarray}
The protocol of quantum key distribution is similar to BB84
\cite{BB84}, the only difference is that three bases are used
instead of two. The advantage of this scheme over BB84, is that
the former is more secure against the eavesdropping.

It has been shown \cite{6state} that the optimal strategy of
individual eavesdropping to the six-state protocol is to attach to
the qubit $S$ a 4-level ancilla in some state $\ket{\chi}$ and to
make the following transformation:
\begin{eqnarray}
\ket{0}\ket{\chi}\longrightarrow\sqrt{F}\ket{0}\ket{A}+\sqrt{1-F}\ket{1}\ket{B},\\
\ket{1}\ket{\chi}\longrightarrow\sqrt{F}\ket{1}\ket{C}+\sqrt{1-F}\ket{0}\ket{D},
\end{eqnarray}
where $F$, as usual, is the channel fidelity for the qubit $S$,
and the four states in the ancilla state space are chosen so that
$\left\{\ket{A},\ket{C}\right\}\perp\ket{B}\perp\ket{D}$, and
$Re\bra{A}\left.C\right>=2-1/F$.

If the optimal AUEM is used for entangling the qubit $S$ to the
ancilla, we obtain from Eq.~(\ref{standard}) for $d=2$:
\begin{eqnarray}
\ket{0}\ket{\Phi^+}\longrightarrow\ket{0}\left[\frac{\alpha+\beta}{\sqrt{2}}\ket{0}\ket{0}
+\frac{\alpha}{\sqrt{2}}\ket{1}\ket{1}\right]\nonumber\\
+\frac{\beta}{\sqrt{2}}\ket{1}\ket{1}\ket{0},\\
\ket{1}\ket{\Phi^+}\longrightarrow\ket{1}\left[\frac{\alpha+\beta}{\sqrt{2}}\ket{1}\ket{1}
+\frac{\alpha}{\sqrt{2}}\ket{0}\ket{0}\right]\nonumber\\
+\frac{\beta}{\sqrt{2}}\ket{0}\ket{0}\ket{1},
\end{eqnarray}
where the kets are written in the $SXY$ order. It is
straightforward to verify that the four states of ancilla
entangled with the states of the qubit $S$ satisfy the demands
imposed on the states $\ket{A}$, $\ket{B}$, $\ket{C}$, $\ket{D}$
for any value of $\varphi$. That is, the optimal AUEM can be used
for optimal individual eavesdropping to the six state protocol of
quantum cryptography. As we have seen, the entangling interaction
can be designed in such way, that it comprise only two qubits of
three, which may be a significant advantage in the practical
applications of the eavesdropping techniques.

\section{Conclusions}

We have given a definition of asymmetric universal entangling
machine for a $d$-level system and have proven its existence. We
have shown that such a machine requires a $d^2$-level ancilla and
have found the transformation producing maximal possible
entanglement for a given degradation of the signal system
fidelity. The obtained machine could also be called ``depolarizing
channel purificator'', since it realizes the most general unitary
transformation of three $d$-level systems acting as a depolarizing
channel in the state space of one of them. Thus, this machine
represents a generic quantum model for studying various universal
quantum processes.

It has turned out that this machine, AUEM, is a generalization to
a wide variety of universal quantum machines suggested in the past
years, comprising the symmetric and asymmetric quantum cloners,
the symmetric quantum entangler, the UNOT gate and the quantum
information distributor. All these devices are particular
realizations of AUEM for a definite decomposition of the ancilla
state space into direct product of state spaces of two $d$-level
systems. This fact suggests that AUEM can be used in the same
applications as the mentioned machines, having at the same time
more degrees of freedom for tailoring the necessary interaction.
Besides, it may find additional applications in quantum
information procession with partially known states and in
eavesdropping to various quantum cryptographic protocols.

The minimal-interaction representation of AUEM shows that for
sufficiently high fidelity the entangling interaction requires a
$d$-level ancilla only, instead of a $d^2$-level ancilla typically
suggested. This result may simplify significantly the development
of realistic universal quantum machines.

From the fundamental point of view, the present paper shows that
the knowledge of quantum state is not necessary for entangling any
quantum system to another system (which is itself in a known
state). Thus, the famous Schr\"odinger's
\textit{gedankenexperiment} can be modified in such a way that
makes the initial state of the cat irrelevant. At the moment of
box opening, the cat will be in a superposition of its initial
state and all its other possible states \cite{life}.

\begin{acknowledgments}
This work was supported by INTAS grant 2001-2097, by the project
"Quantum Imaging" (IST-2000-26019) of the European Union and by
Belarussian Republican Foundation for Fundamental Research.
D.B.H. uses the opportunity to thank Prof. Rohrlich for discussion
on optimality and Prof. Kolobov for hospitality during the stay in
Lille.
\end{acknowledgments}


\begin{thebibliography}{99}
\bibitem{life} Here we consider the states of
alive and dead cat as different states in the state space of a
very complicated but finite system. We do not discuss the problem
of biological life description in the framework of quantum theory.
\end{thebibliography}
\end{document}